\documentclass[twocolumn,aps,prl,superscriptaddress,showpacs]{revtex4-2}

\usepackage{graphicx}
\usepackage{amsmath}
\usepackage{hyperref}
\usepackage{enumitem}
\usepackage{threeparttable}
\usepackage{tabularx}
\newcolumntype{Y}{>{\centering\arraybackslash}X}
\usepackage{multirow}
\usepackage{booktabs}

\begin{document}

\title{{\fontfamily{qcr}\selectfont
QE-CONVERSE}:  An open-source package for the Quantum ESPRESSO distribution to compute non-perturbatively orbital magnetization from first principles, including NMR chemical shifts and EPR parameters}

\author{S. Fioccola}
\email{sfioccola@laas.fr}
\affiliation{LAAS-CNRS, Université de Toulouse, CNRS, Toulouse, France}

\author{L. Giacomazzi}
\affiliation{CNR - Istituto Officina dei Materiali (IOM), c/o SISSA Via Bonomea 265, IT-34136 Trieste, Italy}

\author{D. Ceresoli}
\affiliation{CNR-SCITEC – Istituto di Scienze e Tecnologie Chimiche “G. Natta”, National Research Council of Italy, via C. Golgi 19, Milano 20133, Italy}

\author{N. Richard}
\affiliation{CEA, DAM, DIF, Arpajon, France}

\author{A. Hemeryck}
\affiliation{LAAS-CNRS, Université de Toulouse, CNRS, Toulouse, France}

\author{L. Martin-Samos}
\affiliation{CNR - Istituto Officina dei Materiali (IOM), c/o SISSA Via Bonomea 265, IT-34136 Trieste, Italy}

\date{\today}

\begin{abstract}
Orbital magnetization, a key property arising from the orbital motion of electrons, plays a crucial role in determining the magnetic behavior of molecules and solids. Despite its straightforward calculation in finite systems, the computation in periodic systems poses challenges due to the ill-defined position operator and surface current contributions. The modern theory of orbital magnetization, formulated in the Wannier representation and implemented within the Density Functional Theory (DFT) framework, offers an accurate solution through the "converse approach." In this paper, we introduce {\fontfamily{qcr}\selectfont QE-CONVERSE}, a refactored and modular implementation of the converse method, designed to replace the outdated routines from Quantum ESPRESSO (version 3.2). {\fontfamily{qcr}\selectfont QE-CONVERSE} integrates recent advancements in computational libraries, including scaLAPACK and ELPA, to enhance scalability and computational efficiency, particularly for large supercell calculations. While {\fontfamily{qcr}\selectfont QE-CONVERSE} incorporates these improvements for scalability, the main focus of this work is provide the community with a performing and accurate first principles orbital magnetization package to compute properties such as Electron Paramagnetic Resonance (EPR) g-tensors and Nuclear Magnetic Resonance (NMR) chemical shifts, specially in systems where perturbative methods fail. We demonstrate the effectiveness of {\fontfamily{qcr}\selectfont QE-CONVERSE} through several benchmark cases, including the NMR chemical shift of ${}^{27}$Al\ in alumina and ${}^{17}$O and ${}^{29}$Si in $\alpha$-quartz, as well as the EPR g-tensor of $_{ }^{n}\Sigma(n\geq 2)$ radicals and substitutional nitrogen defects in silicon. In all cases, the results show excellent agreement with theoretical and experimental data, with significant improvements in accuracy for EPR calculations over the linear response approach.
The {\fontfamily{qcr}\selectfont QE-CONVERSE} package, fully compatible with the latest Quantum ESPRESSO versions, opens new possibilities for studying complex materials with enhanced precision.
\end{abstract}

\maketitle

\section{Introduction}
Orbital magnetization, a fundamental property arising from the orbital motion of electrons, plays a crucial role in determining the magnetic behavior of molecules and solids. Orbital magnetization arises from the breaking of time-reversal (TR) symmetry by spin-orbit (SO) coupling. This phenomenon can occur spontaneously in ferromagnetic materials or can be triggered in nonmagnetic materials by external perturbations, such as magnetic fields. While the calculation of the orbital magnetization in finite systems is straightforward, it becomes challenging in periodic systems due to the ill-defined position operator and the contribution of itinerant surface current \cite{Resta_2010,PhysRevB.74.024408}. Over the past two decades, this challenge has been tackled through the formulation of the modern theory of orbital magnetization \cite{PhysRevLett.95.137204,PhysRevLett.95.137205} in the Wannier representation (similarly to the modern theory of polarization  using the Berry Phase approach \cite{PhysRevB.47.1651, RevModPhys.66.899}).
This formulation \cite{PhysRevB.74.024408}, enables the calculation of orbital magnetization in the thermodynamic limit within the framework of Density Functional Theory (DFT) in the so-called "the converse approach" \cite{PhysRevB.81.060409, Thon2009}.

Through first-principles orbital magnetization, several derived macroscopic properties can also be computed, such as the Electron Paramagnetic Resonance (EPR) $g$-tensor parameter and Nuclear Magnetic Resonance (NMR) chemical shifts \cite{PhysRevB.81.060409, Thon2009}. 
In systems where spin-orbit coupling cannot be described as a perturbation or the $g$-tensor exhibits a large deviation from the free electron value \cite{PhysRevB.81.060409,PhysRevB.81.195208},  the  "converse" approach is known to be more accurate than the widely exploited linear response (LR) approach \cite{PhysRevB.81.060409}. An example is given by the PbF family of molecules, which due to the degenerate $e$ level occupied by unpaired electrons and the absence of SO coupling along the bond direction, a failure of the LR approach is observed in contrast to the converse approach (see Table III of ref. \cite{PhysRevB.81.060409}).

The converse approach was originally implemented \cite{PhysRevB.81.060409} in the {\fontfamily{qcr}\selectfont PWscf} code of the, now, obsolete Quantum Espresso (QE) version 3.2. In this paper, we present a deep refactoring and revised implementation of the former converse routines gathered in a stand-alone module called {\fontfamily{qcr}\selectfont QE-CONVERSE} package. {\fontfamily{qcr}\selectfont QE-CONVERSE} is fully compatible with latest Quantum ESPRESSO versions \cite{Giannozzi2017} (from version 7.2 onward). It fully exploits recent advances and improvements in QE libraries (mpi scalability including additional parallelization levels and use of optimized linear algebra libraries such as scaLAPACK \cite{slug} or ELPA \cite{Marek_2014}) that enhance computational performance in larger systems, compared to the former version (see Supplementary Materials). However, the main goal of the present work is not focused on computational efficiency, but rather on ensuring the high accuracy of the calculated properties, such as EPR g-tensors and NMR chemical shifts, which are often crucial in the study of complex systems. With respect to the original version, few bugs have been solved, in particular, the calculation of orbital magnetization, NMR chemical shifts and EPR g-tensor parameters for any cell symmetry.

This paper is organized as follows. In Section 2, we briefly review the theoretical background of the converse method. Sections 3 and 4 cover the implementation, installation, and usage of the {\fontfamily{qcr}\selectfont QE-CONVERSE} package.  Finally, Section 5 and 6 showcase the new {\fontfamily{qcr}\selectfont QE-CONVERSE} version on selected becnhmark systems going from molecules to extended systems. 

\section{Theoretical background}
In the following, we only outline the theoretical background of quantities derived from the calculation of orbital magnetization, i.e EPR g-tensor and NMR shielding tensor. For a detailed derivation of the orbital magnetization within the modern theory, we refere the reader to Refs. \cite{Resta_2010, PhysRevB.74.024408, PhysRevLett.95.137204, PhysRevLett.95.137205}.
\subsection{Converse EPR}
The converse method for calculating the EPR $g$-tensor, introduced in Ref \cite{PhysRevB.81.060409}, can be summarized as follows. The starting point is an independent particles Kohn-Sham Hamiltonian that preserves the translation symmetry of the crystal but breaks Time-Reversal (TR) symmetry. In the all-electron (AE) formalism, the Hamiltonian in atomic units is given by:

\begin{equation}
H_{\rm{AE}}=\frac{1}{2}\Big[\mathbf{p}+\alpha A(\mathbf{r})\Big]^{2}+V(\mathbf{r})+\frac{\alpha^{2}g'}{8}\sigma\cdot\Big[\nabla V(\mathbf{r})\times \mathbf{p}+\alpha A(\mathbf{r})\Big]
\label{eq:1}
\end{equation}
where $A(\mathbf{r})$ is the symmetric gauge vector potential equal to $\frac{1}{2} \mathbf{B}\times\mathbf{r}$, $\alpha=\frac{1}{c}$ is the fine structure constant, $g'=2(g_{e}-1)$  and $\sigma$ are the Pauli matrices. In contrast to the linear response approach, the converse method bypasses the perturbation of the Hamiltonian: the spin-orbit coupling (SO) term is explicitly incorporated into the self-consistent field calculations, while the spin other orbit (SOO) term, that in general gives a small contribution to the $g$-tensor, is neglected. The orbital magnetization is formally given by the Hellmann-Feynman equations as: 
\begin{equation}
\mathbf{M}=f_{n}\sum\limits_{n} \langle \psi_{n}|-\frac{\partial H_{\rm{AE}}}{\partial \mathbf{B}}|\psi_{n}  \rangle
\label{eq:2}
\end{equation}
where $f_{n}$ is the occupation of the eigenstate $n$ where the expectation values is taken on ground-state spinor $\psi_{n}$.
Equation (\ref{eq:2}) provides a direct evaluation of orbital magnetization in finite systems. However, extending this to periodic systems is challenging due to the undefined nature of the position operator. To overcome this limitation, the equation has been reformulated in the Wannier representation, which allows for the computation of orbital magnetization in the thermodynamic limit using the Berry-phase formula:
\begin{equation}
\mathbf{M}=-\frac{\alpha N_{c}}{2N_{k}} \rm{Im}\sum\limits_{n,\mathbf{k}} f_{n,\mathbf{k}}  \langle \partial_{\mathbf{k}} u_{n,\mathbf{k}}|\times \left ( H_{\mathbf{k}}+\epsilon_{n,\mathbf{k}} -2\epsilon_{F} \right ) |  \partial_{\mathbf{k}} u_{n,\mathbf{k}} \rangle
\label{eq:3}
\end{equation} 
where $H_{\mathbf{k}}$ is the crystal Hamiltonian with $\mathbf{B}=0$, $\epsilon_{n,\mathbf{k}}$ and $u_{n,\mathbf{k}}$ are its eigenvalues and eigenvectors, $\epsilon_{F}$ is the Fermi level, $N_{c}$ is the number of cells and $N_{k}$ is the number of $\mathbf{k}$ points. The $ |  \partial_{k} u_{n,\mathbf{k}} \rangle$ is $\mathbf{k}$ the derivative of the Bloch wave function for $n$ occupied states. 
In the context of an AE method, Equation (\ref{eq:3}) is valid for both normal periodic insulators and metals, as well as Chern insulators with a non-null Chern invariant. In the norm-conserving pseudopotential framework, instead, where the core regions of AE wave functions are replaced by smoother pseudo waves (PS), the gauge including projection augmented wave (GIPAW) reconstruction is needed, as shown by Pickard and Mauri's theory \cite{PhysRevB.63.245101}. The GIPAW approach transforms the Hamiltonian of Equation (\ref{eq:1}) into pseudo GIPAW Hamiltonian, that at zero order of the magnetic field yields to the following terms:
\begin{equation}
H_{\rm{GIPAW}}^{(0,0)}=\frac{1}{2}\mathbf{p}^{2}+V_{loc}(\mathbf{r})+\sum_{\mathbf{R}}^{}V_{\mathbf{R}}^{NL}
\label{eq:4}
\end{equation} 
\begin{equation}
H_{\rm{GIPAW}}^{(1,0)}=\frac{g'}{8}\alpha ^{2}\left [ \mathbf{\sigma } \cdot (\nabla V_{loc}(\mathbf{r})\times \mathbf{p})+\sum_{\mathbf{R}}^{}F_{\mathbf{R}}^{NL} \right ]
\label{eq:5}
\end{equation} 
where $V_{loc}(\mathbf{r}) $ is the local Kohn-Sham potential and $ V_{\mathbf{R}}^{NL} $ is the nonlocal pseudopotential in the separable Kleinmann-Bylander (KB) form: 
\begin{equation}
V_{\mathbf{R}}^{NL}=\sum_{nm}^{}| \,\beta_{\mathbf{R},n}  \rangle v_{\mathbf{R},nm} \langle \beta_{\mathbf{R},m}|
\label{eq:6}
\end{equation} 
while $F_{\mathbf{R}}^{NL}$ is the separable nonlocal operator accounting the so-called paramagnetic contribution of the atomic site $\mathbf{R}$, that given the set of GIPAW projectors $|\, \widetilde{\rho}_{\mathbf{R},n} \rangle$ can be written as:\\
\begin{equation} 
F_{\mathbf{R}}^{NL}=\sum_{\mathbf{R},nm}^{}| \,\widetilde{\rho}_{\mathbf{R},n}  \rangle \sigma \cdot {f}_{\mathbf{R},nm} \langle \, \widetilde{\rho}_{\mathbf{R},m} | \,
\label{eq:7}
\end{equation} 
The expression ${f}_{\mathbf{R},nm}$  refer to the paramagnetic GIPAW integral and is given by Eq. 10 of Ref. \cite{PhysRevLett.88.086403}. At the first order in the magnetic field, the GIPAW transformation yields the two terms:
\begin{equation}
H_{\rm{GIPAW}}^{(0,1)}=\frac{\alpha}{2}\mathbf{B}\cdot \left(\mathbf{L}+\sum_{\mathbf{R}}^{}\mathbf{R}\times \frac{1}{i}\left [ \mathbf{r}, {V}_\mathbf{R}^{NL} \right ]\right)
\label{eq:8}
\end{equation} 
\begin{align}
H_{\rm{GIPAW}}^{(1,1)} &= \frac{g'}{16}\alpha^{3}\mathbf{B} \cdot \left(\mathbf{r}\times(\sigma\times \nabla V_{loc}) + \sum_{\mathbf{R}}^{}E_{\mathbf{R}}^{NL} \right. \nonumber \\
&\quad \left. + \sum_{\mathbf{R}}\mathbf{R}\times \frac{1}{i}\left[ \mathbf{r}, {F}_\mathbf{R}^{NL} \right] \right)
\label{eq:9}
\end{align} 
where $E_{\mathbf{R}}^{NL}$ similar to $F_{\mathbf{R}}^{NL}$, is the separable nonlocal operator accounting for the diamagnetic contribution which uses the GIPAW integral ${e}_{\mathbf{R},nm}$ given by Eq. 11 of Ref.\cite{PhysRevLett.88.086403}: 
\begin{equation}
E_{\mathbf{R}}^{NL}=\sum_{\mathbf{R},nm}^{}| \, \widetilde{\rho}_{\mathbf{R},n}  \rangle \cdot {e}_{\mathbf{R},nm} \langle \widetilde{\rho}_{\mathbf{R},m}|
\label{eq:10}
\end{equation} 
Inserting the Hamiltonian first order terms, $H_{\rm{GIPAW}}^{(0,1)}$+$H_{\rm{GIPAW}}^{(1,1)}$, into Eq. (\ref{eq:2}) we obtain the final expression of orbital magnetization in the presence of nonlocal pseudopotential: 
\begin{equation}
\mathbf{M} = \mathbf{M}_{bare}+\Delta \mathbf{M}_{NL}+ \Delta \mathbf{M}_{para}+ \Delta \mathbf{M}_{dia}
\label{eq:11}
\end{equation} 
\begin{equation}
\mathbf{M}_{bare}=\frac{\alpha}{2}\sum_{\mathbf{R}}^{}\langle\mathbf{r}\times\frac{1}{i} \left[\mathbf{r},  H_{\rm{GIPAW}}^{(0,0)}+H_{\rm{GIPAW}}^{(1,0)}  \right]      \rangle
\label{eq:12}
\end{equation} 
\begin{equation}
\Delta \mathbf{M}_{NL}=\frac{\alpha}{2}\sum_{\mathbf{R}}^{}\langle{(\mathbf{R-r})}\times\frac{1}{i} \left[\mathbf{r-R}, {V}_\mathbf{R}^{NL}   \right]      \rangle
\label{eq:13}
\end{equation} 
\begin{equation}
\Delta \mathbf{M}_{para}=\frac{g'\alpha^3}{16}\sum_{\mathbf{R}}^{}\langle{(\mathbf{R-r})}\times\frac{1}{i} \left[\mathbf{r-R}, {F}_\mathbf{R}^{NL}   \right]      \rangle
\label{eq:14}
\end{equation} 
\begin{equation}
\Delta \mathbf{M}_{dia}=\frac{g'\alpha^3}{16}\sum_{\mathbf{R}}^{}\langle{} {E}_\mathbf{R}^{NL}       \rangle
\label{eq:15}
\end{equation} 
where $\langle{}...\rangle$ stands for $\sum\limits_{n,\mathbf{k}} f_{n,\mathbf{k}}\langle \psi_{n}|...|\psi_{n}\rangle$. The $\mathbf{M}_{bare}$ can be easily computed by the modern theory of orbital magnetization of Eq. (\ref{eq:3}) where $H_{\mathbf{k}}$ is the is the GIPAW Hamiltonian, and $\epsilon_{n,\mathbf{k}}$ and $u_{n,\mathbf{k}}$ are its eigenvalues and eigenvectors. The $\Delta \mathbf{M}_{NL}$, $\Delta \mathbf{M}_{para}$ and $\Delta \mathbf{M}_{dia}$ are respectively the nonlocal, paramagnetic and diamagnetic correction and their expression in the framework of Bloch functions can be found in Appendix B of Ref. \cite{PhysRevB.81.184424}. In the converse approach, for the sake of simplicity, the orbital magnetization is computed using a collinear approach where the total spin is aligned along an "easy axis" denoted as $\mathbf{e}$, which is typically one of three magnetic directions. The choice of $\mathbf{e}$ affects the spin-orbit coupling within the system. Since the spin-orbit coupling is dependent on the orientation of the spin axis $\mathbf{e}$, the computed orbital magnetization is a function of $\mathbf{e}$. Therefore, to obtain a comprehensive understanding of the orbital magnetization within the system, it is necessary to compute the orbital magnetization as a function of the three different orientations of $\mathbf{e}$.
Finally, from the orbital magnetization the deviation of the $g$-tensor $\Delta g_{\mu\nu}$ from the free electron values ($g_{e}=2.002 \ 319$) is obtained by the variation in  $\mathbf{M}$ with a spin flip: \\
\begin{equation}
\Delta g_{\mu\nu}=-\frac{2}{\alpha}\mathbf{e_{\mu}}\cdot \frac{\mathbf{M}(\mathbf{e_{\nu}})-\mathbf{M}(\mathbf{-e_{\nu}})}{S-(-S)}=-\frac{2}{\alpha S} \mathbf{e_{\mu}}\cdot \mathbf{M}(\mathbf{e_{\nu}})
\label{eq:16}
\end{equation}
where $\mu\nu$ are Cartesian directions of the magnetic field, and $S$ is the total spin. 
\subsection{Converse NMR}
The formulation of the converse method for the calculation of NMR chemical shielding is very similar to the formulation used for EPR calculations, with the main difference being the inclusion of an additional vector potential term. This term corresponds to a magnetic dipole $\mathbf{m}_{s}$ centered at the atom $\mathbf{s}$ and coordinates $\mathbf{r}_{s}$: 
\begin{equation}
A_{s}(\mathbf{r})=\frac{\mathbf{m}_{s}\times(\mathbf{r}-\mathbf{r}_{s})}{|\mathbf{r}-\mathbf{r}_{s}|^3}
\label{eq:17}
\end{equation} 
The inclusion of this vector potential modifies the AE Hamiltonian to:
\begin{equation}
H_{\rm{AE}}=\frac{1}{2}\Bigg\{\mathbf{p}+\alpha\Big[ A(\mathbf{r})+A_{s}(\mathbf{r})\Big]\Bigg\}^{2}+V(\mathbf{r})
\label{eq:18}
\end{equation}
When applying the GIPAW transformation—similar to the method used for EPR calculations—and considering the zeroth order in the external magnetic field, $H_{\rm{GIPAW}}^{(0,0)}$ matches Equation (\ref{eq:4}). At the first order, the term $H_{\rm{GIPAW}}^{(1,0)}$ is expressed as follows:
\begin{equation}
H_{\rm{GIPAW}}^{(1,0)}=\frac{\alpha}{2} \Bigg[ \mathbf{p}\cdot A_{s}(\mathbf{r}) +A_{s}(\mathbf{r})\cdot\mathbf{p} \Bigg]+\sum_{\mathbf{R}}^{}K_{\mathbf{R}}^{NL}
\label{eq:19}
\end{equation} 
where $K_{\mathbf{R}}^{NL}$ addressing the paramagnetic contribution of the magnetic dipole and takes the form of a nonlocal operator: 
\begin{equation}
K_{\mathbf{R}}^{NL}=\frac{\alpha}{2} \sum_{\mathbf{R},nm}^{}| \, \widetilde{\rho}_{\mathbf{R},n}  \rangle  \mathbf {k}_{\mathbf{R},nm} \langle \widetilde{\rho}_{\mathbf{R},m} |
\label{eq:20}
\end{equation} 
where ${k}_{\mathbf{R},nm}$ refers to the GIPAW paramagnetic integrals that take into account the effect of the additional vector potential. These integrals, obtained from a set of AE partial waves $|\phi_{\mathbf{R},n}\rangle$ and pseudopotential partial waves $|\,\widetilde{\phi}_{\mathbf{R},n}\rangle$, can be expressed as:
\begin{align}
k_{\mathbf{R},nm} &=  \langle\phi_{\mathbf{R},n}| \, \mathbf{p}\cdot A_{s}(\mathbf{r}) +A_{s}(\mathbf{r})\cdot \mathbf{p}\,|\,\phi_{\mathbf{R},n}\rangle \nonumber \\
&\quad -\langle\widetilde{\phi}_{\mathbf{R},n}|\,\mathbf{p}\cdot A_{s}(\mathbf{r}) +A_{s}(\mathbf{r})\cdot \mathbf{p} \,| \, \widetilde{\phi}_{\mathbf{R},n}\rangle
\label{eq:21}
\end{align} 
At first order in the external magnetic field, we obtain a term $H_{\rm{GIPAW}}^{(0,1)}$ that is equal to Eq. (\ref{eq:8}), while the term $H_{\rm{GIPAW}}^{(1,1)}$ becomes: 
\begin{align}
H_{\rm GIPAW}^{(1,1)} &= \frac{\alpha}{2}\mathbf{B} \cdot \Bigg(\mathbf{r}\times A_{s}(\mathbf{r}) 
  + \sum_{\mathbf{R}} J_{\mathbf{R}}^{\rm NL} \nonumber \\
  &\quad + \sum_{\mathbf{R}} \mathbf{R} \times \frac{1}{i}\left[ \mathbf{r}, {K}_\mathbf{R}^{\rm NL} \right] \Bigg)
  \label{eq:22}
\end{align} 
where $J_{\mathbf{R}}^{\rm NL}$ is the nonlocal operator: 
\begin{equation}
J_{\mathbf{R}}^{NL}=\sum_{\mathbf{R},nm}^{}|\,\widetilde{\rho}_{\mathbf{R},n}  \rangle  \mathbf {j}_{\mathbf{R},nm} \langle \widetilde{\rho}_{\mathbf{R},m} |
\label{eq:23}
\end{equation} 
and the $\mathbf {j}_{\mathbf{R},nm}$ its GIPAW integrals addressing the diamagnetic contribution:
\begin{align}
j_{\mathbf{R},nm} &=  \langle\phi_{\mathbf{R},n}|\,(\mathbf{r-R})\times A_{s}(\mathbf{r}) |\,\phi_{\mathbf{R},n}\rangle \nonumber \\
&\quad - \langle\widetilde{\phi}_{\mathbf{R},n}|\,(\mathbf{r-R})\times A_{s}(\mathbf{r}) |\,\widetilde{\phi}_{\mathbf{R},n}\rangle
\label{eq:24}
\end{align} 
As in the case of EPR, solving the Hellmann-Feynman equation using the first-order in magnetic field Hamiltonian terms, yields the orbital magnetization as the sum of $\mathbf{M}_{bare}$, $\Delta \mathbf{M}_{NL}$, $\Delta \mathbf{M}_{para}$ and $\Delta \mathbf{M}_{dia}$ terms like the Eq. (\ref{eq:11}). However, in the calculation of NMR, the $\Delta \mathbf{M}_{para}$ and $\Delta \mathbf{M}_{dia}$ terms are defined by replacing the nonlocal operators $E_{\mathbf{R}}^{NL}$ and $F_{\mathbf{R}}^{NL}$, respectively, with $K_{\mathbf{R}}^{NL}$ and $J_{\mathbf{R}}^{NL}$: 
\begin{equation}
\Delta \mathbf{M}_{para}=\frac{\alpha}{2}\sum_{\mathbf{R}}^{}\langle{(\mathbf{R-r})}\times\frac{1}{i} \left[\mathbf{r-R}, {K}_\mathbf{R}^{NL}   \right]      \rangle
\label{eq:25}
\end{equation}  
\begin{equation}
\Delta \mathbf{M}_{dia}=\frac{\alpha}{2}\sum_{\mathbf{R}}^{}\langle{} {J}_\mathbf{R}^{NL}       \rangle
\label{eq:26}
\end{equation} 
Finally, the chemical shielding tensor $\sigma_{s,\alpha\beta}\ $ is obtained from the derivative of the orbital magnetization $\mathbf{M}$ with respect to a magnetic point dipole $\mathbf{m}_{s} $, located at the site of atom $s$:
\begin{equation}
\sigma_{s,\alpha\beta}\ = \delta_{\alpha\beta}-\Omega\frac{\partial \mathbf{M}_{\beta}}{\partial \mathbf{m}_{s}}
\label{eq:27}
\end{equation} 
where $\delta_{\alpha\beta}$ is the Kronecker delta and $\Omega$ is the volume of the simulation cell.
\section{Implementation}
\begin{figure*} [h!]
\centering
\includegraphics[scale=0.8]{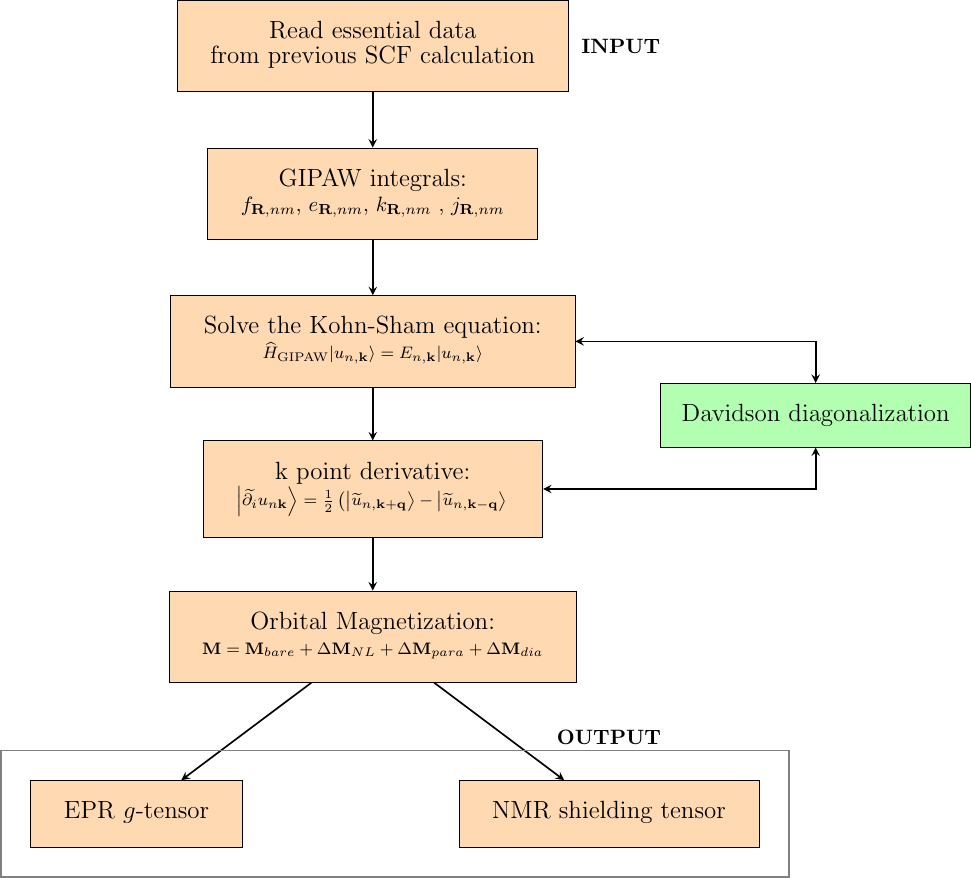}
\caption{Schematic flowchart of the {\fontfamily{qcr}\selectfont QE-CONVERSE} code. }
\label{fig:flowchart}
\end{figure*}

The initial stage of the {\fontfamily{qcr}\selectfont QE-CONVERSE} code involves the application of specialized norm-conserving pseudopotentials with GIPAW reconstruction. These pseudopotentials, apart from the standard norm-conserving pseudopotentials, integrate a complete set of AE core wavefunctions as well as AE ($\phi_{\mathbf{R},n}$) and PS ($\widetilde{\phi}_{\mathbf{R},n}$) partial waves. During this initial phase, the code uses these pseudopotentials to calculate the necessary GIPAW integrals—namely, (${f}_{\mathbf{R},nm}$, ${e}_{\mathbf{R},nm}$, ${k}_{\mathbf{R},nm}$ and ${j}_{\mathbf{R},nm}$) and to determine the GIPAW projectors ($\widetilde{\rho}_{\mathbf{R},n}$). Details on constructing these pseudopotentials are provided in Ref.\cite{PhysRevB.81.184424}, and a non-exhaustive library of GIPAW pseudopotentials is available in Ref.\cite{pseudoGIPAW}.  
(Figure \ref{fig:flowchart} illustrates the overall workflow of the {\fontfamily{qcr}\selectfont
QE-CONVERSE} code.)

\begin{figure*}[h!]
\centering
\includegraphics[scale=0.82]{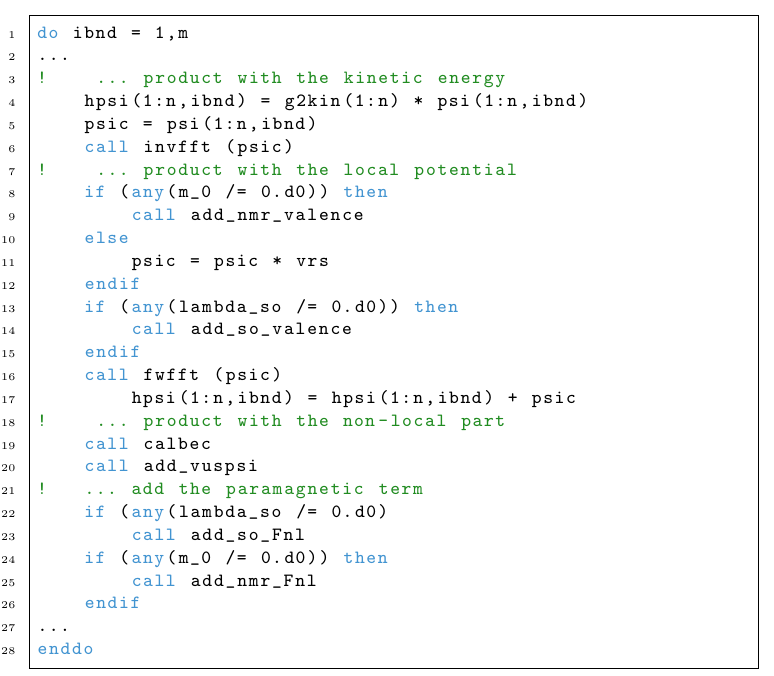}
\caption{Pseudo code sections of \texttt{h\_psi\_gipaw} routine of {\fontfamily{qcr}\selectfont
QE-CONVERSE} code.}
\label{fig:figure1}
\end{figure*}

The first step in the process involves solving the Kohn-Sham equation self-consistently, where the Hamiltonian is the sum of GIPAW Hamiltonian terms at zero order magnetic field ($H_{\rm{GIPAW}}^{(0,0)}$+$H_{\rm{GIPAW}}^{(1,0)}$). In the {\fontfamily{qcr}\selectfont QE-CONVERSE} code, these terms are constructed by the \texttt{h\_psi\_gipaw} routine, which, similar to \texttt{h\_psi} of the {\fontfamily{qcr}\selectfont PWscf} code, computes the product of the kinetic energy, the local part of the potential, and its non-local part in KB form of the Hamiltonian matrix with wavefunctions. 
The input flag, {\fontfamily{qcr}\selectfont lambda\textunderscore so $\neq$ 0}  (see Appendix B),
activates the EPR $g$-tensor calculations, adding the SO term and its paramagnetic contribution $F_{\mathbf{R}}^{NL}$ to the Hamiltonian. While in the NMR chemical shift calculation, if {\fontfamily{qcr}\selectfont m\textunderscore 0 $\neq$ 0}, the vector potential and its paramagnetic term $K_{\mathbf{R}}^{NL}$ are added. A schematic representation of \texttt{h\_psi\_gipaw} routine is presented in Figure \ref{fig:figure1}.
Once a new ground state is reached, the next step is the calculation of the $\mathbf{k}$ derivative of the Bloch wave functions implementing the covariant finite difference formula. This formula calculates the finite difference between gauge-invariant "dual" states in the mesh direction $i$ according to:
\begin{equation}
|\widetilde{\partial _{i}}u_{n \mathbf{k}}\rangle=\frac{1}{2}\left ( |\,\widetilde u_{n, \mathbf{k+q}}\rangle - |\,\widetilde u_{n, \mathbf{k-q}}\rangle  \right )  
\label{eq:28}
\end{equation}
Here, $|\,\widetilde u_{n, \mathbf{k+q}}\rangle$ is constructed as a linear combination of the occupied state $|u_{n, \mathbf{k+q}}\rangle$ at neighboring mesh point $\mathbf{q}$: 
\begin{equation}
|\,\widetilde u_{n, \mathbf{k+q}}\rangle=\sum_{n'}^{}\left ( S_{\mathbf{k+q}}^{-1} \right )_{n'n} | u_{n', \mathbf{k+q}}\rangle 
\label{eq:29}
\end{equation}
where $(S_{\mathbf{k+q}}^{} )_{nn'}$ is the overlap matrix defined as:
\begin{equation} 
    (S_{\mathbf{k+q}}^{} )_{nn'}=\langle u_{n, \mathbf{k}}| u_{n', \mathbf{k+q}}\rangle
    \label{eq:30}
\end{equation}
The {\fontfamily{qcr}\selectfont QE-CONVERSE} code runs the covariant finite procedure as follow:

\begin{enumerate}[label=\arabic*.]
    \item Loop over $\mathbf{k}$-points.
    \begin{enumerate}[label=\arabic*.,left=1em]
        \item Read $|u_{n,\mathbf{k}} \rangle$ from the disk, all bands.
        \item Loop over $\mathbf{q}$-points: $\pm q \hat{x}$, $\pm q \hat{y}$, $\pm q \hat{z}$.
        \begin{enumerate}[label=\arabic*.,left=0.5em]
            \item Loop over bands
            \begin{enumerate}[label=\arabic*.,left=0.5em]
                \item Diagonalize the KS Hamiltonian at $\mathbf{k}$+$\mathbf{q}$
                \item Compute the overlap matrix: $(S_{\mathbf{k+q}}^{} )_{nn'}$
                \item Invert the matrix: $  ( S_{\mathbf{k+q}}^{-1}  )_{n'n}$ 
                \item Compute the dual state: $|\,\widetilde u_{n, \mathbf{k+q}}\rangle$
                \item Compute the finite difference:
                $\frac{1}{2}\left ( |\,\widetilde u_{n, \mathbf{k+q}}\rangle - |\,\widetilde u_{n, \mathbf{k-q}}\rangle  \right )$
            \end{enumerate}
        \end{enumerate}
    \end{enumerate}
    
    \item Save the derivative on disk.
\end{enumerate} 
The diagonalization of the Hamiltonian, undertaken by both the routines dedicated to SCF calculations and those executing the $\mathbf{k}$ derivative of the Bloch functions, employs the Davidson method, addressed by the {\fontfamily{qcr}\selectfont cegterg} routine of Quantum ESPRESSO (Fig.\ref{fig:flowchart}).

\begin{figure*}[h!]
\centering
\includegraphics[scale=0.95]{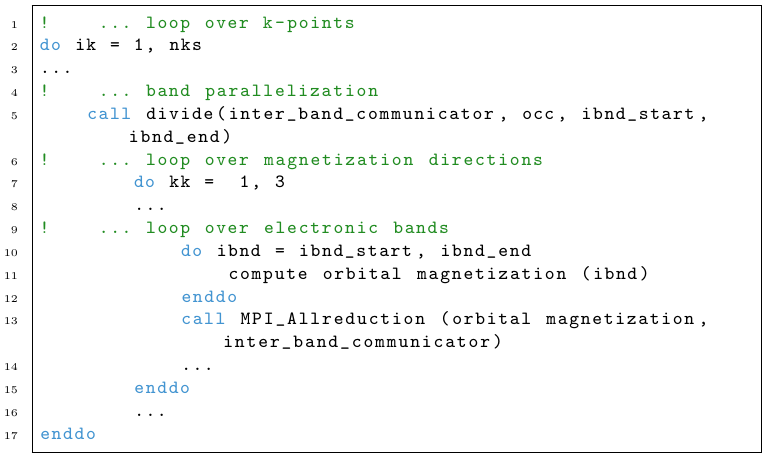}
\caption{Pseudo code sections of orbital magnetization calculation routine.}
\label{fig:figure2}
\end{figure*}

In the outline of the code, the third step is the orbital magnetization calculation implementing the Eq. (\ref{eq:3}). $\mathbf{M}_{bare}$ is computed using a nested loop over k-points, magnetic directions, and electronic bands. This step involves calculating the expectation value of the Hamiltonian operator within the $|\widetilde{\partial _{i}}u_{n \mathbf{k}}\rangle$ state (Fig.\ref{fig:figure2}). In order to speed up the calculation of this part and improve the scalability when running the calculation at high parallel cores, in the {\fontfamily{qcr}\selectfont QE-CONVERSE} code, with respect to the former version of converse code, we add a new level of parallelization which is based on band group bands and electronic bands MPI communicator, by slitting the \texttt{ibnd} index across processors (Fig.\ref{fig:figure2}). 
The nonlocal, paramagnetic, and diamagnetic correction terms are then computed starting from the converged ground state and using the nonlocal operators $E_{\mathbf{R}}^{NL}$, $F_{\mathbf{R}}^{NL}$, and $K_{\mathbf{R}}^{NL}$, $J_{\mathbf{R}}^{NL}$ respectively for the EPR $g$-tensor and for the NMR chemical shielding. All the magnetization terms are summed as specified by Eq. (\ref{eq:11}) to yield the total orbital magnetization. At the end, once $\mathbf{M}$ is obtained, the deviation $\Delta g_{\mu\nu}$ is calculated according to Eq. (\ref{eq:16}). Alternatively, in the case of NMR, it is derived with respect to the magnetic dipole according to Eq. (\ref{eq:27}). 

Appendix A (Table \ref{table:benchmark}{}) shows the comparisons between PbF radicals g-tensors in Ref. \cite{PhysRevB.81.060409} and the present converse implementation. 
\section{Installation and usage}
The repository of our code is released at GitHub (\url{https://github.com/mammasmias/QE-CONVERSE.git}.
A Fortran 90 compiler and the Quantum ESPRESSO package must be previously installed. To take advantage of the enhancements in linear algebra operations, the scaLAPACK package or the ELPA library are required.
To install the code, the program files must be copied or cloned from the repository. Then, the code must be configured with the desired version of QE by typing {\fontfamily{qcr}\selectfont\$./configure --with-qe-source="QE folder containing make.inc"} replacing {\fontfamily{qcr}\selectfont"QE folder containing make.inc"} with the path to the main directory of QE. Finally, by typing {\fontfamily{qcr}\selectfont \$ make} the executable binary {\fontfamily{qcr}\selectfont qe-converse.x} is compiled in {\fontfamily{qcr}\selectfont /bin/} directory. To execute a converse calculation with the {\fontfamily{qcr}\selectfont QE-CONVERSE}, the first step is to perform a SCF calculation using the {\fontfamily{qcr}\selectfont PWscf} code of Quantum ESPRESSO on a physically meaningful (e.g. a geometrically optimized) structure. For an EPR g-tensor calculation, the SCF calculation must be spin-polarized; this is can be archived by including the flag {\fontfamily{qcr}\selectfont nspin= 2} in the input for {\fontfamily{qcr}\selectfont PWscf} code. The $\mathbf{k}$-points grid mesh should be set up without applying symmetry operations to the crystal, and given that the Kohn-Sham Hamiltonian breaks time-reversal symmetry, it is necessary to assume that the $k$ points and -$k$ points are not equivalent. This grid construction can be easily implemented including the flags {\fontfamily{qcr}\selectfont nosym=.true.} and {\fontfamily{qcr}\selectfont noinv=.true.}in the input for the SCF calculation. In the same directory where the calculation with the {\fontfamily{qcr}\selectfont PW} program is executed, the {\fontfamily{qcr}\selectfont QE-CONVERSE} code is launched with {\fontfamily{qcr}\selectfont ./qe-converse.x}  using an input file (an example for EPR and NMR calculation is provided in Appendix B). It is essential that the 
{\fontfamily{qcr}\selectfont prefix} and {\fontfamily{qcr}\selectfont outdir} flags match those used in the SCF calculation with the {\fontfamily{qcr}\selectfont PWscf} program. This is because the {\fontfamily{qcr}\selectfont QE-CONVERSE} code reads essential data such as atomic coordinates, ground state function, pseudopotentials, and the $\mathbf{k}$-points grid from the XML file in the {\fontfamily{qcr}\selectfont outdir} directory of a preceding spin-polarized SCF calculation. The flag {\fontfamily{qcr}\selectfont q} in the input file refers to the small vector used into covariant derivative finite difference formula. Since its value influences the accuracy of the derivative, it is suggested to keep its default value of 0.01. To compute the EPR $g$-tensor, the flag {\fontfamily{qcr}\selectfont lambda\textunderscore so(1,..,3)} must be included in the input file to incorporate the SO term into the Hamiltonian. The integer value within the parentheses indicates the direction in which the orbital magnetization is calculated. When calculating the NMR chemical shift instead, it is necessary to introduce into input file the nuclear dipole moment flag, {\fontfamily{qcr}\selectfont m\textunderscore0(1,..,3)} and the index of the atom carrying the dipole moment using the designated flag {\fontfamily{qcr}\selectfont m\textunderscore0\textunderscore atom}. 
Appendix B describes briefly the most relevant textual inputs and outputs.

\section{EPR $g$-tensor benchmarks} 
The input and textual output files can be found in the {\fontfamily{qcr}\selectfont /examples/} directory of the official code repository.

\subsection{$_{ }^{n}\Sigma(n\geq 2)$ molecules} 
In this first example, we illustrate the EPR $g$-tensor of $_{ }^{2}\Sigma$ and high-spin $_{ }^{3}\Sigma$ ($ \text{O}_2$ and $\text{S} \text{O}$) radicals. We performed the calculations in a cubic  cell with a large volume of 8000 $\text{\AA}^3$ and the Brillouin zone sampled at $\Gamma$ point with an energy cutoff of 100 Ry where the Kohn-Sham equation is solved using the PBE functional. The following Table \ref{table:radicals} compares the $\Delta g$ components in ppm from the {\fontfamily{qcr}\selectfont QE-CONVERSE} code with available experimental data and other computational approaches, such as Gauge-Independent Atomic Orbital–Density Functional Theory (DFT-GIAO) \cite{doi:10.1021/jp963060t, doi:10.1021/jp010457a} and coupled-cluster singles and doubles level of theory (CCSD) \cite{10.1063/1.4979680}. Table \ref{table:radicals} demonstrates that the results from {\fontfamily{qcr}\selectfont QE-CONVERSE} are well aligned with the other computational models and are in good agreement with experimental observations. It is important to note that for some of the radicals, a direct comparison with experimental $g$-value is not fully suitable due the use of the noble-gas matrix isolation technique, which can affect the value of EPR parameters \cite{10.1063/1.5130174}. 

\newcolumntype{Y}{>{\centering\arraybackslash}X}

\begin{table}[h]
    \centering
    \begin{threeparttable}
        \caption{The principal components of $\Delta g$ (ppm) calculated in this
        work (TW) for \texorpdfstring{$_{ }^{n}\Sigma(n\geq 2)$}{n-Sigma (n >= 2)} molecules compared with other theoretical approaches and experimental results.}
        \setlength{\tabcolsep}{4pt}
        \begin{tabularx}{0.9\linewidth}{clYYYY}
            \hline \hline
            Molecule &  & DFT-GIAO\tnote{a}  & CCSD\tnote{b}  & TW & Expt.  \\ 
            \midrule
            \multirow{3}{*}{$\text{H}_2\text{O}^+$} 
            & $\Delta g_{xx}$ & 103 & -250 & -219 & 200\tnote{c} \\
            & $\Delta g_{yy}$ & 13824 & 15156 & 12702 & 18800\tnote{c} \\
            & $\Delta g_{zz}$ & 5126 & 4800 & 4485 & 4800\tnote{c} \\
            \midrule
            \multirow{3}{*}{$\text{N} \text{O}_2$} 
            & $\Delta g_{xx}$ & 4158 & 3327 & 4850 & 3300\tnote{d} \\
            & $\Delta g_{yy}$ & -13717 & -10792 & -14319 & -10300\tnote{d} \\
            & $\Delta g_{zz}$ & -760 & -630 & -795 & 700\tnote{d} \\
            \midrule
            \multirow{2}{*}{$\text{Si} \text{H}_3$} 
            & $\Delta g_{{\perp}}$ & 2402 & 1833 & 2183 & 5000\tnote{e} \\
            & $\Delta g_{{\parallel}}$ & -105 & -133 & -102 & 1000\tnote{e} \\
            \midrule
            \multirow{2}{*}{$\text{C} \text{O}^+$} 
            & $\Delta g_{{\perp}}$ & -3129 & -2436 & -3316 & -3200\tnote{d} \\
            & $\Delta g_{{\parallel}}$ & -138 & -162 & -143 & -1400\tnote{d} \\
            \midrule
            \multirow{2}{*}{$ \text{O}_2$} 
            & $\Delta g_{{\perp}}$ & 3100 &  & 3224 & 2900\tnote{f} \\
            & $\Delta g_{{\parallel}}$ & -300 &  & -450 & \tnote{h} \\
            \midrule
            \multirow{2}{*}{$\text{S} \text{O}$} 
            & $\Delta g_{{\perp}}$ & 4800 & & 4721 & 3600\tnote{g} \\
            & $\Delta g_{{\parallel}}$ & -300 &  & -342 & \tnote{h} \\
            \hline \hline
        \end{tabularx}
        
        \begin{tablenotes}
            \item[a] Ref.\cite{doi:10.1021/jp963060t} for the $_{ }^{2}\Sigma$ molecules and Ref.\cite{doi:10.1021/jp010457a} for the high-spin $_{ }^{3}\Sigma$ molecules computed with RPBE functional. 
            \item[b] Ref.\cite{10.1063/1.4979680}.
            \item[c] Gas phase results derived from spin rotation data (Ref.\cite{Knight}).
            \item[d] Experiment in Neon gas matrix isolation (Ref.\cite{Lushington}).
            \item[e] Experimental values for static radical undergo restricted rotation at 4.2 K in Krypton matrix (Ref.\cite{10.1063/1.1727825}).
            \item[f] Ref.\cite{PhysRev.97.951}
            \item[g] Ref.\cite{10.1246/bcsj.42.886}
            \item[h] The experimental $\Delta g_{{\parallel}}$ value is negligible (Ref. \cite{doi:10.1021/jp010457a}).
        \end{tablenotes}
        \label{table:radicals}
        
    \end{threeparttable}
\end{table}


\subsection{Substitutional Nitrogen in Silicon}
\begin{figure*}[h!]
\centering
\includegraphics[scale=0.5]{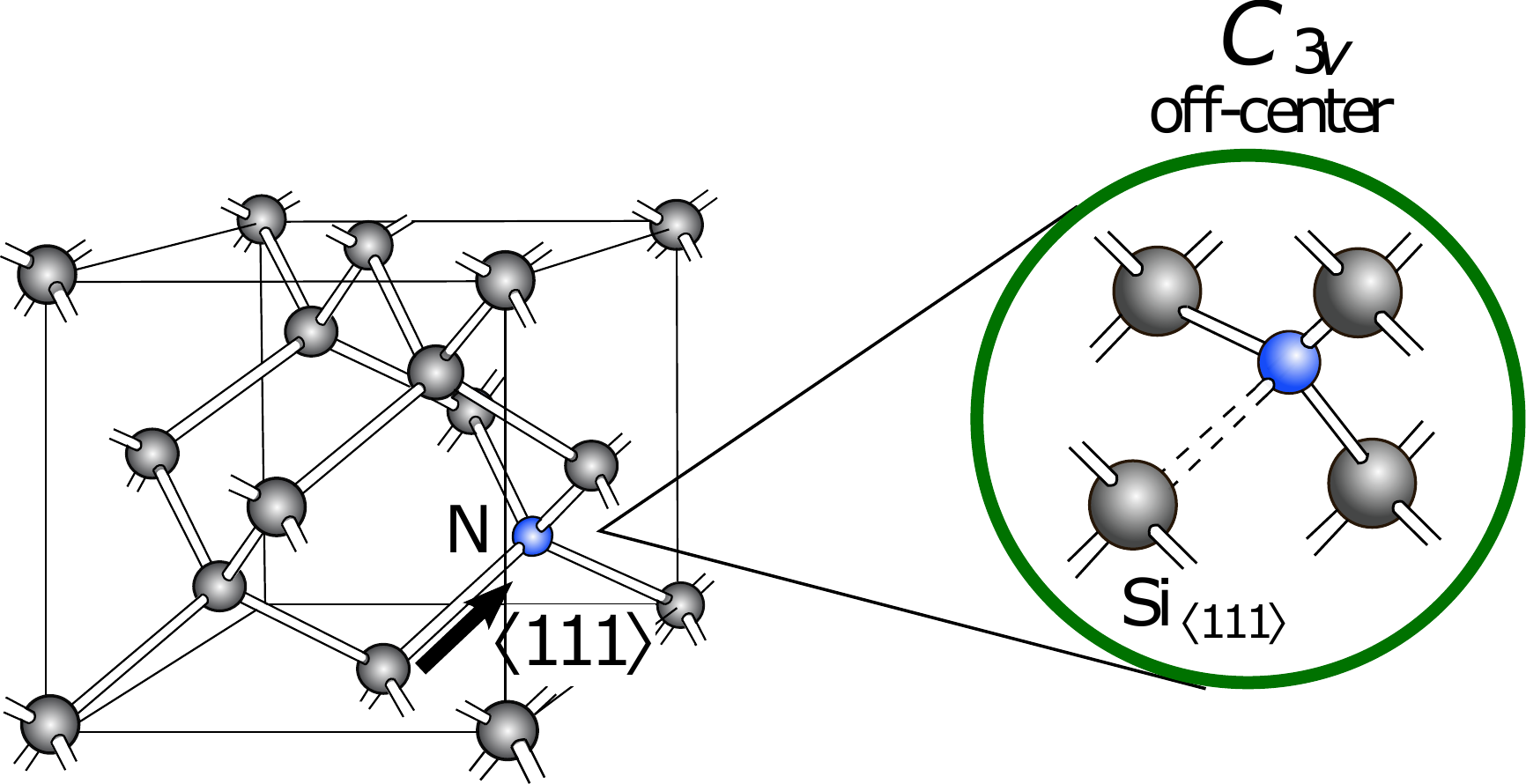}
\caption{Graphic representation of the off-center configuration for (N\textsubscript{Si}) defect. }
\label{fig:figure4}
\end{figure*}

The following section presents the EPR $g$-tensor calculation for the substitutional Nitrogen defect in Silicon (N\textsubscript{Si}). This defect has recently garnered significant theoretical and experimental attention due to its potential role as a spin donor in silicon-based quantum devices \cite{PhysRevB.89.115207,nano14010021}. It is characterized by the presence of two minima, so that the Nitrogen atom transitions from the tetra-coordinated on-center (T\textsubscript{d}) configuration site to a tri-coordinated off-center configuration (with C\textsubscript{3v} symmetry) along one of the ⟨111⟩ crystal directions, through a pseudo Jahn-Teller mechanism \cite{nano13142123}. In this example, we focus solely on the off-center configuration (\ref{fig:figure4}), corresponding to the SL5 experimental EPR signals. Regarding the $g$-tensor in SL5, g\textsubscript{1} corresponds to the component aligned with the principal axis parallel to the ⟨111⟩ crystal direction (or g\textsubscript{$\parallel$}), while g\textsubscript{2} and g\textsubscript{3} represent the isotropic perpendicular components to the ⟨111⟩ crystal direction (or g\textsubscript{$\perp$}).
All ab-initio calculations were performed using the PBE exchange-correlation functional, together with norm-conserving Trouiller-Martins pseudopotentials with GIPAW reconstruction and employing a plane wave basis set with a kinetic energy cutoff of 84 Ry. To model the (N\textsubscript{Si}) off-center point defect we used a 512 atoms supercell in which a substitutional Nitrogen atom is embedded within a silicon cell. Geometry optimization was conducted by sampling at the $\Gamma$ point and setting the ionic convergence threshold to 0.026 eV/\r{A} \cite{nano13142123}. In the subsequent SCF and EPR $g$-tensor calculation, we ensured convergence of the parameters by sampling $\mathbf{k}$-points on a $3\times 3\times 3$ mesh grid without considering symmetry, resulting in the integration of the Brillouin zone over a total of 54 $\mathbf{k}$-points. The SCF convergence threshold was set to $1\cdot 10^{-8}$ Ry.

\begin{table}[]
\setlength{\tabcolsep}{10pt}
\caption{The EPR $g$ tensor for the substitutional Nitrogen 
point defect in silicon (N$_{\rm Si}$) calculated in this 
work using the {\fontfamily{qcr}\selectfont QE-CONVERSE } code compared to the values of  Ref. \cite{nano13142123},
in which a linear response approach is implemented. Results are compared 
to experimental data of Ref. \cite{PhysRevB.89.115207}.}
\vskip0.3cm
\label{table:epr}
\begin{center}
\begin{tabular}{llll}
\hline \hline
 Ref.   & $g_{1}$ &  $g_{2}$ & $g_{3}$\\ 
\hline
 This work & 2.00209     & 2.00860     & 2.00860     \\ 
 Ref. \cite{nano13142123}      & 2.00200     & 2.00890     & 2.00780     \\ 
 Expt. \cite{PhysRevB.89.115207}      & 2.00219     & 2.00847     & 2.00847     \\ 
 \hline \hline
\end{tabular}
\end{center}
\end{table}

The Table \ref{table:epr} presents the values of the $g$-tensor calculated using the {\fontfamily{qcr}\selectfont QE-CONVERSE} code, compared with experimental results \cite{PhysRevB.89.115207} and the $g$-tensor calculated by Simha et al. \cite{nano14010021} where the linear response method was implemented \cite{PhysRevLett.88.086403}. Our results demonstrate good agreement with the experimental data. Moreover, our calculations yield a $g$-tensor that is fully consistent with the axial symmetry along the ⟨111⟩ crystal directions, as shown by experiments, thus improving upon the results previously computed by the linear response method of Ref. \cite{nano14010021}.
\section{NMR chemical shifts benchmarks}
\subsection{\(\textbf{${}^{17}$O}\) and \(\textbf{${}^{29}$Si}\) NMR chemical shifts in $\alpha$-quartz}\ \ 
In the first example, we predicted the ${}^{17}$O and ${}^{29}$Si NMR chemical shifts in $\alpha$-quartz using the converse method through the {\fontfamily{qcr}\selectfont QE-CONVERSE} code. We first computed the isotropic shielding tensor, $\sigma_{iso}$, for both isotopes and then extrapolated their chemical shifts using the relationship $\delta_{iso} = \sigma_{ref} - \sigma_{iso}$, where we used reference values, $\sigma_{ref}$, of 324.51 ppm and 253.34 ppm for the ${}^{29}$Si and ${}^{17}$O nuclei, respectively \cite{PhysRevB.81.184424}. We performed the DFT calculations with the PBE functional, using a 70 Ry kinetic energy cut-off and integrating the Brillouin zone on a $2\times 2\times 2$ mesh grid. We present the results of our calculations in the following Table \ref{tab:quartz}, where the results are compared with experimental observations. Notably, the calculations demonstrate convergence and alignment with experimental findings.
\begin{table}[h]
\centering
\begin{threeparttable}
\caption{ Calculated and experimental ${}^{29}$Si and ${}^{17}$O NMR chemical shift of  $\alpha$-quartz.}
\setlength{\tabcolsep}{10pt} 
\begin{tabular}{lll}
\hline\hline
            & \multicolumn{2}{c}{\(\delta_{iso}\) (\(\mathrm{ppm}\))} \\ \cline{2-3} 
structure   & Expt. \tnote{a} & This work \\ \hline
$\alpha$-quartz & \multicolumn{2}{c}{} \\
${}^{29}$Si & -107.73  \tnote & -107.53 \\
${}^{17}$O & \phantom{-0}40.08 \tnote & \phantom{-0}43.09 \\ \hline\hline
\end{tabular}
\begin{tablenotes}
\item[a] Reference \cite{doi:10.1021/ja027124r}
\end{tablenotes}
\label{tab:quartz}
\end{threeparttable}
\end{table}
\subsection{\(\textbf{${}^{27}$Al}\) NMR chemical shifts in alumina}
\begin{figure*}[h!]
  \centering
  \includegraphics[scale=0.07]{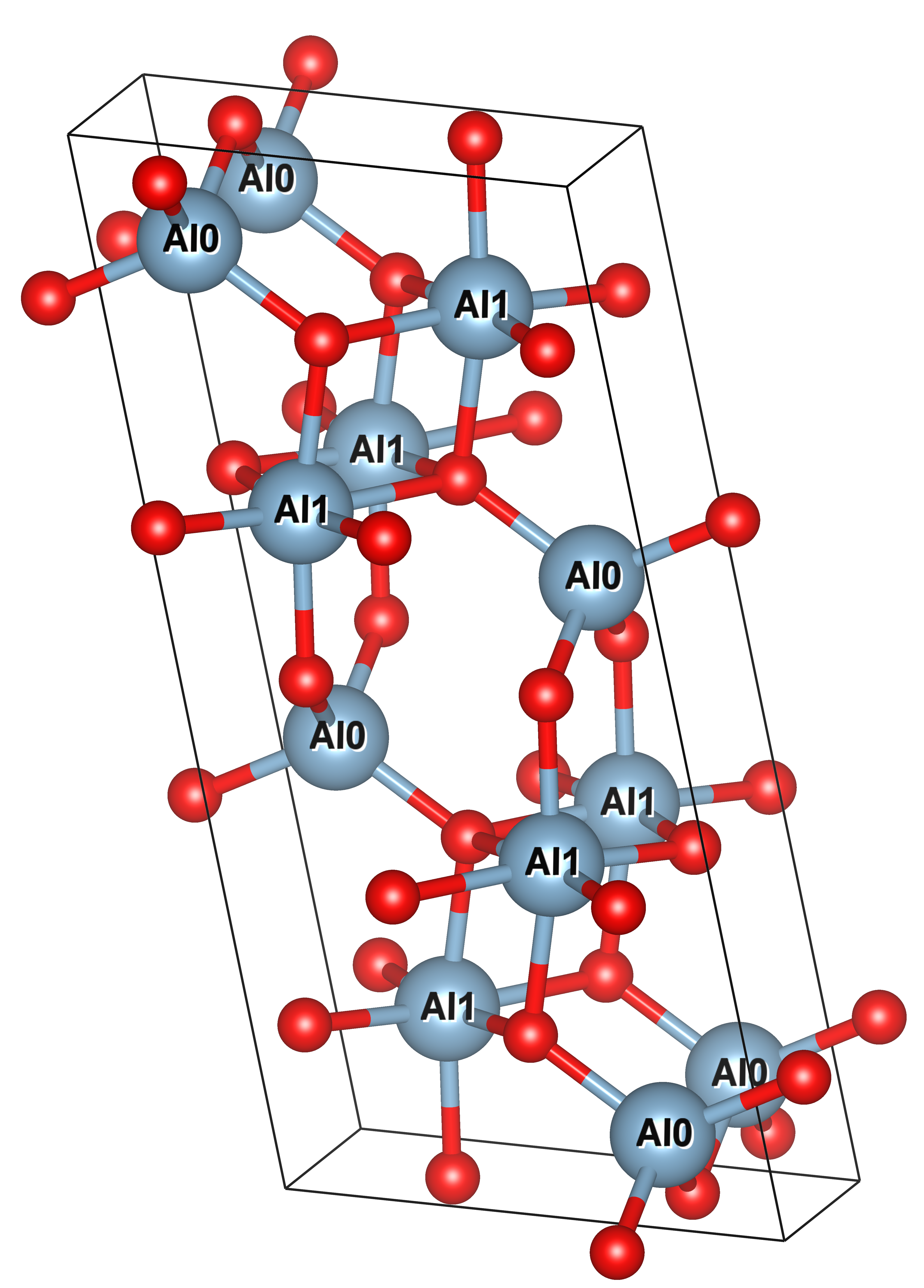}
  \caption{Unit cell of \(\theta\)-alumina. \textbf{Al1} refers to the octahedrally coordinated aluminum atoms (Al\textsubscript{oct}) while \textbf{Al0} are the tetrahedrally coordinated (Al\textsubscript{tet}). }
  \label{fig:figure5}
\end{figure*}

\ 
Corundum (\(\alpha\)-Al\(_2\)O\(_3\)) represents the most thermodynamically stable crystalline form of alumina and it can be synthesized through the calcination process starting from unactivated gibbsite (AlOH)\(_3\), which transforms into corundum via \(\gamma\) and \(\theta\)-Al\(_2\)O\(_3\) transitions at temperatures exceeding 400°C. Determining the structure and the temperature at which transitions phase occur is crucial and it has applications in various industrial fields. For example, \(\gamma\)-Al\(_2\)O\(_3\) is used as a catalyst in the petrochemical industry. The most useful technique for this purpose is solid-state NMR of ${}^{27}$Al of which the detailed interpretation of the experimental spectra can be enhanced by first-principle calculations. In this study, we apply the {\fontfamily{qcr}\selectfont QE-CONVERSE} code to perform ab-initio calculations of the NMR chemical shift of \ ${}^{27}$Al\ in \(\theta\) phase of alumina (Figure \ref{fig:figure5}). In this phase, the experimental spectra of ${}^{27}$Al\ exhibit two distinct signals: one corresponding to the aluminum atoms octahedrally coordinated (Al\textsubscript{oct}), and the other to those tetrahedrally coordinated (Al\textsubscript{tet}) by oxygen atoms \cite{ODELL2007169}. Implementing the Eq. (\ref{eq:27}) we computed the absolute chemical shift tensors of \ ${}^{27}$Al\ nuclei that we converted into isotropic chemical shieldings using $\sigma_{iso}=Tr\left [ \sigma_{s}/3 \right ]$. Finally, we compared this last one with experimental isotropic chemical shifts by using the expression: $\delta _{iso}=\sigma _{ref}-\sigma _{iso}$, where $\sigma _{ref}$ in this work is the isotropic shielding of \ ${}^{27}$Al\ in \(\alpha\) phase, used as reference. \\ We performed DFT calculations with the PBE generalized gradient approximation and using the norm-conserving Trouiller-Martins pseudopotentials with GIPAW reconstruction. We used a cutoff energy of 70 Ry. We first relaxed a 200 atoms \(\theta\)-Al\(_2\)O\(_3\) supercell sampling at the $\Gamma$ point and setting the ionic convergence threshold to 0.026 eV/\r{A}. We then computed the SCF and NMR calculation setting the $\mathbf{k}$-points on a $2\times 2\times 2$ mesh grid without symmetric constraint. We converged the groud-state to $1\cdot 10^{-9}$ Ry. In Table \ref{tab:my-table} below, we present the chemical shifts of \ ${}^{27}$Al\ obtained in this work, compared to experimental \cite{ODELL2007169} and theoretical data from Ref. \cite{PhysRevB.84.235119}, where a linear response method was implemented. Our results show very close agreement with both the experimental and previous theoretical results
\begin{table*}
\end{table*}
\begin{threeparttable}
\caption{Calculated ${}^{27}$Al\ NMR chemical shifts of \(\theta\)-Al\(_2\)O\(_3\) from this work and from reference \cite{ODELL2007169} compared with experimental observations \cite{PhysRevB.84.235119}.}
\setlength{\tabcolsep}{10pt} 
\begin{tabular}{lllll}
\hline\hline
            & \multicolumn{4}{c}{\(\delta_{iso}\) (\(\mathrm{ppm}\))}                           \\ \cline{2-5} 
structure   & \multicolumn{2}{c}{previous} & \multicolumn{2}{c}{this work} \\ \cline{2-3}
Al site     & Expt.             & Th.       & \multicolumn{2}{l}{}          \\ \hline
\(\theta\)-Al\(_2\)O\(_3\) & \multicolumn{4}{c}{}                                         \\
Al\textsubscript{oct}       & -3.0 (1.0)\tnote{a}       & -4.9\tnote{b}      & \multicolumn{2}{c}{-6.3}      \\
Al\textsubscript{tet}       & 66.5 (1.0)\tnote{a}      & 62.6\tnote{b}     & \multicolumn{2}{c}{61.0}      \\ \hline\hline
\end{tabular}
\begin{tablenotes}
\item[a] Reference \cite{ODELL2007169}
\item[b] Reference \cite{PhysRevB.84.235119}
\end{tablenotes}
\label{tab:my-table}
\end{threeparttable}
\section*{Conclusions}  
In this paper, we prensent the {\fontfamily{qcr}\selectfont QE-CONVERSE} code, which implements the converse method—a non-perturbative approach to calculating orbital magnetization and related properties, such as the EPR g-tensor and the NMR chemical shift, from first principles. {\fontfamily{qcr}\selectfont QE-CONVERSE} is a standalone package compatible with Quantum ESPRESSO (version 7.2 and onward), replacing the previously implemented, now outdated, PWscf code from QE 3.2. Compared to the former code, {\fontfamily{qcr}\selectfont QE-CONVERSE} integrates modern linear algebra libraries, such as scaLAPACK and ELPA, which enhance the computational performance of NMR and EPR parameter calculations, especially for large supercell. However, the primary focus of {\fontfamily{qcr}\selectfont QE-CONVERSE} is on improving the accuracy of the calculations, especially for EPR and NMR parameters, where it may provide (depending on the system) more reliable results compared to perturbative methods such as the linear response approach. This enhanced accuracy opens new possibilities for studying complex materials with greater precision and for obtaining more accurate insights into their magnetic properties.
\section*{Acknowledgments} This work has received funding from the NextGenerationEU European initiative through the Italian Ministry of University and Research, PNRR Mission 4, Component 2 - ICSC Italian National Center for High Performance Computing, big data and quantum computing - spoke 7 , project code  HPC (Centro Nazionale 01 CN0000013) CUP B93C22000620006; and by the project SQQS (ab initio control of defects for quantum bits: screening, quantifying, qualifying and selecting)  Defi Cle Region Institut Quantique Occitanie (FRANCE). All the calculations were performed using the HPC resources of CALMIP Mesocenter (Toulouse) with grant P1555. 
L.G. acknowledges  CINECA for computational resources under the project ISCRA-C HP10C2GG7R "NFFAM".
The authors are active members of the Multiscale And Multi-Model ApproacheS for Materials In Applied Science consortium (MAMMASMIAS consortium), and acknowledge the efforts of the consortium in fostering scientific collaboration.

\section*{Conflicts of Interest}
The authors declare no conflict of interest. The funders had no role in the design of the study; in the collection, analyses or interpretation of data; in the writing of the manuscript; or in the decision to publish the results.

\section*{Appendix A : Ref. \cite{PhysRevB.81.060409} {\it{vs}} current implementation}
We benchmarked {\fontfamily{qcr}\selectfont QE-CONVERSE} by comparing the $\Delta g$ for selected diatomic radicals belonging to the PbF family with those obtained from the converse code integrated in {\fontfamily{qcr}\selectfont PWscf} code of version 3.2.  All the results are collected in Table \ref{table:benchmark}. The calculations were 
performed in a cubic repeated cell with a large volume of 8000 $\text{\AA}^3$ and the Brillouin zone sampled at $\Gamma$ point with an energy cutoff of 100 Ry. The Perdew-Burke-Ernzerhof (PBE) exchange-correlation functional was \cite{PhysRevLett.77.3865} employed.
The PbF family radicals have only one electron within the p shell and due to the absence of SO coupling, they exhibit a nearly vanishing $g$-tensor along the bond direction ($\Delta g_{\parallel}\approx -2 \times 10^{6} $ ppm), as predicted by the converse method \cite{PhysRevB.81.060409}. Table \ref{table:benchmark} shows that the results here calculated with the {\fontfamily{qcr}\selectfont QE-CONVERSE} code   
and those previously obtained with the converse method implemented in {\fontfamily{qcr}\selectfont PWscf} 3.2 \cite{PhysRevB.81.060409} well match. \\
\begin{table}[h]
\centering
\setlength{\tabcolsep}{10pt} 
\caption{The $\Delta g$ in ppm for the PbF family radicals calculated in this work using the {\fontfamily{qcr}\selectfont QE-CONVERSE} code compared to the values of {\fontfamily{qcr}\selectfont PWscf} 3.2.}
\label{table:benchmark}
\begin{threeparttable}
\begin{tabular}{clll}
\hline \hline 
\multicolumn{1}{l}{} &          & PWscf 3.2\tnote{a} & QE-CONVERSE \\ \hline 
\multirow{2}{*}{CF}  & $\Delta g_{\parallel}$     & -1999719  & -2000148    \\
                     & $\Delta g_{\perp}$         & -553      & -542        \\  
\multirow{2}{*}{SiF} & $\Delta g_{\parallel}$     & -1995202  & -2000073    \\
                     & $\Delta g_{\perp}$         & -2470     & -2459       \\           
\multirow{2}{*}{GeF} & $\Delta g_{\parallel}$     & -1998078  & -2000050    \\
                     & $\Delta g_{\perp}$         & -39101    & -38534      \\  
\multirow{2}{*}{SnF} & $\Delta g_{\parallel}$     & -1996561  & -2000006    \\
                     & $\Delta g_{\perp}$         & -142687   & -141228     \\  
\multirow{2}{*}{PbF} & $\Delta g_{\parallel}$     & -1999244  & -2000042    \\
                     & $\Delta g_{\perp}$         & -556326   & -552755     \\  
\hline \hline
\end{tabular}
\begin{tablenotes}
\item[a] Reference \cite{PhysRevB.81.060409}
\end{tablenotes}
\end{threeparttable}
\end{table}

\section*{Appendix B: Brief input and output description} 
The input consists on a single fortran namelist \texttt{\&input\textunderscore{}qeconverse} which the keywords are summarised in Table \ref{table:configurations}. \\
Example of input for EPR $g$-tensor calculation. {\fontfamily{qcr}\selectfont
\${DIR}} is an integer from 1 to 3 that denotes the spin-orbit coupling direction. 
\begin{verbatim}
&input_qeconverse
    prefix = 'example_EPR'
    outdir = './scratch/'
    diagonalization = 'david'
    verbosity = 'high'
    q_gipaw = 0.01
    dudk_method = 'covariant'
    diago_thr_init = 1d-4
    conv_threshold = 1e-8
    mixing_beta = 0.5
    lambda_so(${DIR}) = 1.0
/
\end{verbatim}
\vskip 0.5cm
Example of input for NMR calculation. {\fontfamily{qcr}\selectfont
\${DIR}} is an integer from 1 to 3 that referes to nuclear dipole moment direction while {\fontfamily{qcr}\selectfont
\${i}} indicates the atoms that brings the dipole.
\begin{verbatim}
&input_qeconverse
    prefix = 'example_NMR'
    outdir = './scratch/'
    diagonalization = 'david'
    verbosity = 'high'
    q_gipaw = 0.01
    dudk_method = 'covariant'
    diago_thr_init = 1d-4
    conv_threshold = 1e-8
    mixing_beta = 0.5
    m_0(${DIR}) = 1.0
    m_0_atom = ${i}
/
\end{verbatim}
\renewcommand{\arraystretch}{1.5} 
\begin{table*}[t]
\begin{threeparttable}
\caption{Input namelist and its description.}
\begin{tabularx}{\textwidth}{l l l X} 
\hline \hline
\textbf{Keyword} & \textbf{Type} & \textbf{Default} & \textbf{Description} \\ \hline
\textbf{prefix} & character & \texttt{prefix} & Prefix of files saved by program \texttt{pw.x}. The value of this keyword must be the same used in the SCF calculation. \\
\textbf{outdir} & character & \texttt{./} & Temporary directory for \texttt{pw.x} restart files. The value of this keyword must be the same used in the SCF calculation. \\
\textbf{diagonalization} & string & \texttt{david} & Diagonalization method (only allowed values: \texttt{david}). \\
\textbf{verbosity} & string & \texttt{high} & Verbosity level (allowed values: \texttt{low}, \texttt{medium}, \texttt{high}). \\
\textbf{q\_gipaw} & real & 0.01 & The small wave-vector for the covariant finite difference formula. (units: bohr radius$^{-1}$) \\
\textbf{dudk\_method} & string & \texttt{covariant} & k-point derivative method (only allowed values: \texttt{covariant}). \\
\textbf{diag\_thr\_init} & real & 10$^{-7}$ & Convergence threshold (ethr) for iterative diagonalization. (units: Ry$^2$) \\
\textbf{conv\_threshold} & real & 10$^{-8}$ & Convergence threshold for the diagonalization in the SCF step. (units: Ry$^2$) \\
\textbf{mixing\_beta} & real & 0.5 & Mixing factor for self-consistency. \\
\textbf{lambda\_so(1,..,3)} & real & 0.0 & Cartesian components of electron spin. The value (1,..,3) denotes the spin-orbit coupling direction. (units: Bohr magneton) \\
\textbf{m\_0(1,..,3)} & real & 0.0 & Cartesian components of nuclear dipole. The value (1,..,3) denotes the nuclear dipole moment direction. (units: nuclear magneton) \\
\textbf{m\_0\_atom} & integer & 0 & Atom index carrying the nuclear magnetic dipole. \\
\\ \hline \hline
\end{tabularx}
\label{table:configurations}
\end{threeparttable}
\end{table*}
\vskip 0.5cm
The textual output contains the key data are summarised in Table \ref{table:magnetization}.
\begin{table*}[t]
\begin{threeparttable}
\caption{Orbital magnetization and related quantities.}
\begin{tabularx}{\textwidth}{l X} 
\hline \hline
\textbf{Keyword} & \textbf{Description} \\ \hline
\textbf{M\_LC} & \textit{Local Circulation} orbital magnetization in atomic units \cite{PhysRevB.74.024408}. \\
\textbf{M\_IC} & \textit{Itinerant Circulation} orbital magnetization in atomic units \cite{PhysRevB.74.024408}. \\
\textbf{Delta\_M} & Nonlocal, paramagnetic, and diamagnetic GIPAW correction in atomic units to the orbital magnetization. \\
\textbf{M\_total} & Orbital magnetization in atomic units computed as the sum of \textbf{M\_LC}, \textbf{M\_IC}, and \textbf{Delta\_M}. \\
\textbf{delta\_g RMC} & Relativistic mass correction in ppm to the $\Delta g$ \cite{PhysRevLett.88.086403}. \\
\textbf{delta\_g RMC (GIPAW)} & GIPAW correction in ppm to the \textbf{delta\_g RMC}. \\
\textbf{delta\_g SO} & $\Delta g$ in ppm arising from SO coupling. \\
\textbf{delta\_g tot} & $\Delta g$ in ppm computed as the sum of \textbf{delta\_g RMC}, \textbf{delta\_g RMC (GIPAW)}, and \textbf{delta\_g SO}. \\
\textbf{Chemical shift (ppm)} & Chemical shielding tensor in ppm. \\
\textbf{Core shift (ppm)} & Core contribution in ppm to chemical shielding tensor. \\
\\ \hline \hline
\end{tabularx}
\label{table:magnetization}
\end{threeparttable}
\end{table*}


\bibliographystyle{apsrev4-2}
\bibliography{cpc}

\end{document}


\title{{\fontfamily{qcr}\selectfont
QE-CONVERSE}:  An open-source package for the Quantum ESPRESSO distribution to compute non-perturbatively orbital magnetization from first principles, including NMR chemical shifts and EPR parameters}

\author{S. Fioccola}
\email{sfioccola@laas.fr}
\affiliation{LAAS-CNRS, Université de Toulouse, CNRS, Toulouse, France}

\author{L. Giacomazzi}
\affiliation{CNR - Istituto Officina dei Materiali (IOM), c/o SISSA Via Bonomea 265, IT-34136 Trieste, Italy}

\author{D. Ceresoli}
\affiliation{CNR-SCITEC – Istituto di Scienze e Tecnologie Chimiche “G. Natta”, National Research Council of Italy, via C. Golgi 19, Milano 20133, Italy}

\author{N. Richard}
\affiliation{CEA, DAM, DIF, Arpajon, France}

\author{A. Hemeryck}
\affiliation{LAAS-CNRS, Université de Toulouse, CNRS, Toulouse, France}

\author{L. Martin-Samos}
\affiliation{CNR - Istituto Officina dei Materiali (IOM), c/o SISSA Via Bonomea 265, IT-34136 Trieste, Italy}

\date{\today}

\maketitle

\section{Supplementary Materials}
\subsection{Benchmarking Results: Computational Efficiency of QE-CONVERSE}
In this Supplementary Material, we present benchmarking results that demonstrate the computational speed-up of the updated QE-CONVERSE code compared to its earlier version (PW 3.2). These benchmarks focus on the calculation of the EPR g-tensor for the E-center defect in Silicon (VP)\cite{Pfanner2012}, which results are shown in Table \ref{tab:summary_e_center} and \ref{tab:cpu_times}.
\subsection{the EPR g-tensor for the E-center defect in Silicon (VP)}
We modelled the E-center (VP) point defect in Silicon using a supercell of 215 atoms. Ab initio calculations were performed with the PBE exchange-correlation functional, norm-conserving Trouiller-Martins pseudopotentials with GIPAW reconstruction, and a plane-wave basis set with a kinetic energy cutoff of 50 Ry. The calculations runned on 192 CPUs distributed across 4 nodes. Table \ref{tab:summary_e_center} compares the computational performance of the two codes. The results demonstrate that QE-CONVERSE is six times faster than the previous version (PW 3.2), both for calculations involving a single magnetic direction and for the total computation time.
In Table \ref{tab:cpu_times}, we have broken down the computation time for a single magnetic direction into three distinct contributions corresponding to the main steps performed by the codes. The results demonstrate that modern linear algebra libraries significantly enhance the efficiency of the SCF and k-point derivative steps. As expected, these libraries have no impact on the orbital magnetization step.
\begin{table}[h]
    \centering
    \renewcommand{\thetable}{S1}
    \begin{tabular}{ccccccc}
        \hline
        Cell size (atoms) & k-point & Code & \multicolumn{3}{c}{g component} & Total CPU time (hours) \\
         &  &  & g1 & g2 & g3 &  \\
        \hline
        215 & $2\times2\times2$ & PW 3.2 & 1.9997 & 2.0123 & 2.0075 &  60.0 \\
        215 & $2\times2\times2$ & QE-COVERSE & 1.9997 & 2.0123 & 2.0075 &  10.0 \\
        Exp. \cite{Watkins1964} &  &  & 2.0005 & 2.0112 & 2.0096 &  \\
        \hline
    \end{tabular}
    \caption{Computational time benchmarking for the EPR g-tensor calculation of the E-center defect (VP) in Silicon using a 215-atom supercell. The results compare the performance of QE-CONVERSE (with ScaLAPACK enabled) and the previous version, PW 3.2.}
    \label{tab:summary_e_center}
\end{table}

\begin{table}[h]
    \centering
    \renewcommand{\thetable}{S2}
    \begin{tabular}{ccccccc}
        \hline
        Code & SCF (min) & K-point derivative (min) & Orbital magnetization (min) & Total CPU time (min) \\
        \hline
        PW 3.2       & 353.3 & 796.3 & 50.7 & 1200.3 \\
        QE-COVERSE   & 71.1  & 74.5  & 50.7 & 196.3 \\
        \hline
    \end{tabular}
    \caption{Breakdown of computation time for a single magnetic direction during the EPR g-tensor calculation of the E-center defect (VP) in Silicon. The table shows the time contributions from the three main steps: SCF calculations, k-point derivative evaluations, and the orbital magnetization step.}
    \label{tab:cpu_times}
\end{table}

\bibliographystyle{unsrt}
\bibliography{supplementary}